\documentclass{mn2e}
\usepackage{epsfig}

\def\meszaros{M\'{e}sz\'{a}ros}

\begin{document}
\voffset=-0.5 in

\title[Polarization of afterglows from two-component jets]
{Gamma-Ray Bursts: Polarization of Afterglows from Two-Component
Jets}

\author[X. F. Wu et al.]{X. F. Wu,$^{1 \; \star}$ Z. G. Dai,$^{1 \; \star}$ Y. F. Huang,$^{1 \; \star}$
   and T. Lu$^{2}$
\thanks{E-mail: xfwu@nju.edu.cn(XFW); dzg@nju.edu.cn(ZGD); hyf@nju.edu.cn(YFH); tanlu@mail.pmo.ac.cn(TL)}\\
$^1${\sl Department of Astronomy, Nanjing University, Nanjing
210093,  P. R. China} \\
$^2${\sl Purple Mountain Observatory, Chinese Academy of Sciences,
Nanjing 210008, P. R. China } }
\date{Accepted ......  Received ......; in original form ......
      %(MNRAS, 2002, in press)
      }
%\pubyear{2002}

%\shorttitle{X. F. Wu et al.}{Afterglow polarization from two-component jets}
\maketitle

\begin{abstract}
Polarization behaviors of optical afterglows from
two-component gamma-ray burst jets are investigated, assuming
various configurations for the two components.
In most cases, the observed polarization is dominated by the
inner narrow component for a long period. Interestingly, it is
revealed that different assumptions about the lateral expansion
of the jet can lead to different
evolutions of the position angle of polarization.
The observed afterglow light curve and polarization
behaviors of GRB $020813$ can be well explained by the
two-component jet model. Particularly, the model is able to explain the
constancy of the observed position angle in this event, given that the line of
sight is slightly outside the narrow component.
\end{abstract}


\section{Introduction}
Soon after the discovery of gamma-ray burst (GRB) afterglows, it
was expected that radiation from these GRB embers may be polarized
(Loeb \& Perna 1998; Gruzinov \& Waxman 1999; Medvedev \& Loeb
1999). The first direct detection of polarized afterglows from GRB
$990510$ by Covino et al. (1999) and later confirmation by Wijers
et al. (1999) intrigued subsequent investigations of the geometry
of GRB ejecta (Ghisellini \& Lazzati 1999; Sari 1999; Gruzinov
1999).

Polarization originates naturally from some asymmetries when
relativistic electrons produce non-thermal radiation. To model the
observed polarizations of GRB afterglows, there are basically two
kinds of consideration on the violation of symmetry with respect
to the observer. The first class includes the models assuming
ordered magnetic fields. The magnetic fields can be either locally
ordered, which corresponds to the magnetic domain model (Gruzinov
\& Waxman 1999), or even entirely aligned within the ejecta which
is magnetized by the central engine (Granot \& K\"{o}nigl 2003;
Lyutikov, Pariev \& Blandford 2003; Lazzati et al. 2004; Dai
2004). The second class involves beaming effect, where the GRB
ejecta is assumed to be conically collimated. Observers need to be
off-axis, which is natural since a large viewing angle corresponds
to a large chance possibility. The magnetic fields are postulated
to be randomly but an-isotropically distributed behind the shock.
The simplest configuration is two dimensional, with the magnetic
fields randomly distributed in the shock plane (Ghisellini \&
Lazzati 1999; Laing 1980). In further particular advisements the
level of the magnetic field anisotropy has been parameterized
(Gruzinov 1999; Sari 1999; Granot \& K\"{o}nigl 2003). Previous
works revealed that the simple homogeneous jet is confronted with
abrupt variation of polarization angle by $90^{\circ}$, when the
degree of polarization passes through zero. No intermediate values
for the position angle are expected in this model. In fact, the
observed position angle of some afterglows, e.g. GRBs $990712$
(Rol et al. 2000), $020405$ (Covino et al. 2003a) and $020813$
(Gorosabel et al. 2004), did not exhibit such significant
changes\footnotemark\footnotetext{Significant evolution of
position angles does have been observed in two events, i.e. GRBs
$021004$ and $030329$ (Rol et al. 2003; Greiner et al. 2003).
However, this does not necessarily imply the homogeneous jet
structure since their light curves are peculiar. As first pointed
out by Granot \& K\"{o}nigl (2003), variable afterglow light
curves are expected to be accompanied by variable polarization
light curves. Nakar \& Oren (2004) fitted the light curve and
polarization curve of GRB $021004$ within the angular
inhomogeneous ``patchy" shell model, while Bj\"{o}rnsson,
Gudmundsson $\&$ J\'{o}hannesson (2004) interpreted this afterglow
within the context of the refreshed shock model.}. As indicated by
simulations of jet propagating in the collapsar, the emergent jet
from the stellar envelope is structured (Zhang, Wooley \& Heger
2004). Two approximations on the jet structure had been made
previously. One is the power law structure (\meszaros, Rees \&
Wijers 1998; Dai \& Gou 2001; Rossi, Lazzati \& Rees 2002), while
the other is the Gaussian profile (Zhang \& \meszaros 2002).
Recently, Rossi et al. (2004) had detailedly investigated the
polarization behavior of both a power law structured jet and a
Gaussian jet. The advantage of these kinds of structured jets is
that the position angle keeps constant with time.

In this paper we explore the polarization of GRB afterglows by
assuming a different jet structure, i.e. the two-component jet
structure. The two-component jet model was proposed to explain
observations, such as unusual radio to X-ray spectrum and late
time bumps in some afterglows (Frail et al. 2000; Berger et al.
2003; Huang et al. 2004). Liang \& Dai (2004) found that there
exist two peaks in the histogram of the spectral peak energy
distribution derived from $\nu F_{\nu}$ in GRBs, which also
tentatively implies a two-component structure of GRB jets.  In
Section 2 we describe the dynamics and radiation mechanism of
two-component jets. The afterglow polarization behavior of
two-component jets is presented in Section 3. Observed light curve
and polarization curve of GRB $020813$ are fitted within this
model in Section 3.3. We conclude and discuss our results in
Section 4. In the Appendix, the polarization degree of synchrotron
radiation in an ordered magnetic field as a function of frequency
is given.

\section{Jet hydrodynamics and synchrotron radiation}
For simplicity, we make several assumptions on the hydrodynamic
evolution of a two-component jet. First, the two-component jet
evolves adiabatically. Second, we apply the uniform thin shell
approximation to both components of the jet. Third, we assume that
there is no interaction between the narrow and wide components.
Each component evolves independently and therefore can be treated
separately (see also Huang et al. 2004). Lastly, two different
assumptions are made on the lateral expansion of the jet. One is
to assume that the jet expands laterally at the speed of light,
and the other is to postulate that the jet experiences no lateral
expansion. The former corresponds to the case of a clear-cut
lateral boundary between the uniform jet and the environment,
which was assumed for simplicity in previous works (Rhoads 1999;
Sari, Piran, $\&$ Halpern 1999). However, the physical parameters
near the actual lateral boundary should vary gradually (Zhang et
al. 2004). Recent hydrodynamical calculations showed that for
smooth jet profiles the lateral expansion speed is much less than
the sound speed in the jet co-moving frame even during the
relativistic stage (Kumar $\&$ Granot 2003; Granot $\&$ Kumar
2003). So, whether the jet expands laterally or not is still an
open question. We therefore also discuss the second possibility
that the jet has no sideways expansion. In the following, we
denote the physical parameters of the narrow and wide components
by the subscripts ``\emph{N}" and ``\emph{W}", respectively. The
narrow (wide) component can be described by the isotropic kinetic
energy $E_{N,iso}$ ($E_{W,iso}$), the bulk Lorentz factor
$\gamma_{N}$ ($\gamma_{W}$), and the half-opening angle
$\theta_{N}$ ($\theta_{W}$).

The hydrodynamic evolution can be evaluated in the same way for both components of the jet. One
caveat should be made about the inner boundary of the wide component. The wide component is assumed
to be a hollow cone, with the inner half-opening angle to be constant and the same as the initial
aperture of the narrow component during the whole relativistic stage. However, the hydrodynamics of
the wide component should be hardly affected by this fact. We let $\theta_{0}$ be the initial
half-opening angle of an arbitrary component. After a short coasting phase or reverse-forward shock
interaction epoch, the jet begins to decelerate in the interstellar medium (ISM), with its radius
$R$, bulk Lorentz factor $\gamma$ and half-opening angle $\theta_j$ evolve as (Sari, Piran, $\&$
Narayan 1998)
\begin{equation}
R=5.8\times10^{17}E_{53}^{1/4}n^{-1/4}t_d^{1/4}(\frac{1+z}{2})^{-1/4}\,\,\rm{cm,}
\end{equation}
\begin{equation}
\gamma=11E_{53}^{1/8}n^{-1/8}t_d^{-3/8}(\frac{1+z}{2})^{3/8},\phantom{ssss}
\theta_j\approx\theta_0,
\end{equation}
where the isotropic kinetic energy $E=10^{53}E_{53}$ erg, the ISM
number density $n$ is in units of cm$^{-3}$, $t_{d}$ is the
observer's time in days, and $z$ is the redshift. During this
stage, the hydrodynamics of the jet is the same as that of an
isotropic fireball, which obeys the self-similar solution of
Blandford-McKee (1976). As the jet continues to decelerate, there
is a critical moment when the Lorentz factor
$\gamma=\theta_0^{-1}$, which corresponds to the time
\begin{equation}
t_j=0.645(1+z)E_{53}^{1/3}n^{-1/3}\theta_{0,-1}^{8/3}\,\rm{day},
\end{equation}
where $\theta_0=0.1\theta_{0,-1}$. In the case of no lateral
expansion, the jet hydrodynamics after this time is unchanged and
still can be described by equations (1) and (2). On the other
hand, if the jet expands laterally at the speed of light, the jet
will experience a runway sideways expansion (Rhoads 1999; Sari et
al. 1999). The hydrodynamic quantities in this case when $t>t_j$
evolve as
\begin{equation}
R=6.18\times10^{17}E_{53}^{1/3}n^{-1/3}\theta_{0,-1}^{2/3}\,\rm{cm,}
\end{equation}
\begin{equation}
\gamma=\theta_{0}^{-1}(\frac{t}{t_j})^{-1/2},\phantom{ssssssss}\theta_j\approx\gamma^{-1}.
\end{equation}
Polarized afterglows have been observed usually within
several days (Covino et al. 2002; Bj\"{o}rnsson 2002). In this
paper, we focus on the stage when the jet is still relativistic. The
moment when the jet becomes non-relativistic is typically much
later, unless the ISM density is very
large and comparable to that of giant molecular clouds in the
Galaxy (Dai \& Lu 1999).

The post-shock electrons and magnetic fields are usually assumed
to share the fractions $\epsilon_{e}$ and $\epsilon_{B}$ of the
total internal energy, respectively. Therefore, the magnetic field
in the co-moving frame is
\begin{equation}
B'=\sqrt{32\pi\epsilon_{B} m_{p}n}\gamma c
\approx0.4\gamma\epsilon_{B}^{1/2}n^{1/2}\,\,\rm{G},
\end{equation}
while the minimum Lorentz factor of accelerated electrons is
(Huang et al. 2000)
\begin{equation}
\gamma_{m}=\epsilon_{e}\frac{p-2}{p-1}\frac{m_{p}}{m_{e}}(\gamma-1)+1,
\end{equation}
where $m_{p}$ and $m_{e}$ are the proton and electron masses, and
$p$ is the index of shock-accelerated electron distribution. The
electrons can be cooled down within the dynamical time scale. The
cooling Lorentz factor of electrons by purely synchrotron cooling
is (Sari et al. 1998)
\begin{equation}
\gamma_{c}^{syn}=\frac{6\pi m_{e}c(1+z)}{\sigma_{\rm{T}}\gamma
B'^{2}t},
\end{equation}
where $c$ is the speed of light and $\sigma_{\rm{T}}$ is the
Thomson cross section. Further cooling of electrons by
inverse-Compton (IC) up-scattering of the primary synchrotron
photons can reduce the above cooling Lorentz factor (Waxman 1997;
Wei \& Lu 1998; Panaitescu \& Kumar 2000). Following Sari \& Esin
(2001), we get the final cooling Lorentz factor
\begin{equation}
\gamma_{c}=\frac{\sqrt{1+4\displaystyle\frac{\epsilon_{e}}{\epsilon_{B}}}-1}{2\displaystyle\frac{\epsilon_{e}}{\epsilon_{B}}}\gamma_{c}^{syn}\approx
     \left\{
\begin{array}{l}
\gamma_{c}^{syn},\phantom{ssssss}  \epsilon_{e}\ll\epsilon_{B}, \\
\sqrt{\displaystyle\frac{\epsilon_{B}}{\epsilon_{e}}}\gamma_{c}^{syn},\phantom{s}  \epsilon_{e}\gg\epsilon_{B},\\
 \end{array} \right.
\end{equation}
in the fast-cooling phase ($\gamma_{c}<\gamma_{m}$), and
\begin{equation}
\gamma_{c}\approx\left\{ \begin{array}{l}
\displaystyle(\sqrt{\frac{\epsilon_{B}}{\epsilon_{e}}}\frac{\gamma_{c}^{syn}}{\gamma_{m}})^{2/(4-p)}\gamma_{m},
\phantom{s}  \gamma_{c}^{syn}\leq(\frac{\epsilon_{e}}{\epsilon_{B}})^{1/(p-2)}\gamma_{m}, \\
\displaystyle\gamma_{c}^{syn},  \phantom{ssssssssssssssssss} \gamma_{c}^{syn}\geq(\frac{\epsilon_{e}}{\epsilon_{B}})^{1/(p-2)}\gamma_{m},\\
 \end{array} \right.
\end{equation}
in the slow-cooling phase ($\gamma_{m}<\gamma_{c}$). In deriving
the above $\gamma_{c}$ we adopt $2<p<4$, which seems to be easily
satisfied by both the theoretical prediction and observations (Wu,
Dai, \& Liang 2004). It can be seen from equations (9) and (10)
that $\gamma_{c}$ always equals to $\gamma_{c}^{syn}$ for
$\epsilon_{e}\leq\epsilon_{B}$. In the case of
$\epsilon_{e}>\epsilon_{B}$, the evaluation of $\gamma_{c}$ is
considered in three evolutionary stages. Equation (9) describes
the first IC-dominated fast-cooling case, while equation (10)
represents the following IC-dominated and then
synchrotron-dominated slow-cooling cases. The two characteristic
frequencies of synchrotron radiation in the co-moving frame are
the typical frequency $\nu_{m}'=\displaystyle\frac{eB'}{2\pi
m_{e}c}\gamma_{m}^{2}$ and the cooling frequency
$\nu_{c}'=\displaystyle\frac{eB'}{2\pi m_{e}c}\gamma_{c}^{2}$,
which corresponds to the electrons with minimum Lorentz factor
$\gamma_{m}$ and with cooling Lorentz factor $\gamma_{c}$,
respectively. Sari et al. (1998) showed that the co-moving peak
spectral power per electron is independent of the electron energy,
i.e.
$P'_{\nu',max}=\displaystyle\frac{m_{e}c^{2}\sigma_{\rm{T}}}{3e}B'$,
where $e$ is the electron charge. The co-moving peak intensity is
therefore (Rossi et al. 2004)
\begin{equation}
I'_{\nu',max}=\frac{P'_{\nu',max}}{4\pi}\frac{nR}{3}=\frac{m_{e}c^{2}\sigma_{\rm{T}}}{36\pi
e}B'nR.
\end{equation}

The spectral energy distribution of synchrotron radiation can be
constructed by the characteristic frequencies $\nu_{c}'$,
$\nu_{m}'$ and the peak intensity $I'_{\nu',max}$. In the
fast-cooling phase ($\nu_{c}'<\nu_{m}'$), we have
\begin{equation}
I'_{\nu'}=I'_{\nu',max}\times\left\{ \begin{array}{l}
\displaystyle(\frac{\nu'}{\nu'_{c}})^{1/3}, \phantom{sssssssssssss} \nu'<\nu'_{c},\\
\displaystyle(\frac{\nu'}{\nu'_{c}})^{-1/2}, \phantom{sssssssssss}\, \nu'_{c}<\nu'<\nu'_{m},\\
\displaystyle(\frac{\nu'_{m}}{\nu'_{c}})^{-1/2}(\frac{\nu'}{\nu'_{m}})^{-p/2}, \phantom{s}\, \nu'_{m}<\nu',\\
 \end{array} \right.
\end{equation}
and in the slow-cooling phase ($\nu_{m}'<\nu_{c}'$), we have
\begin{equation}
I'_{\nu'}=I'_{\nu',max}\times\left\{ \begin{array}{l}
\displaystyle(\frac{\nu'}{\nu'_{m}})^{1/3}, \phantom{sssssssssssss} \nu'<\nu'_{m},\\
\displaystyle(\frac{\nu'}{\nu'_{m}})^{-(p-1)/2}, \phantom{ssssssss} \nu'_{m}<\nu'<\nu'_{c},\\
\displaystyle(\frac{\nu'_{c}}{\nu'_{m}})^{-(p-1)/2}(\frac{\nu'}{\nu'_{c}})^{-p/2}, \nu'_{c}<\nu'.\\
 \end{array} \right.
\end{equation}
We do not include the synchrotron-self-absorption effect which may
be important mainly at the radio wavelength. Since afterglow
polarizations had been detected mostly in optical bands, we
will only investigate the temporal evolution of optical polarization.

\section{Polarization}
\subsection{Formulation}
\begin{figure}
\epsfig{figure=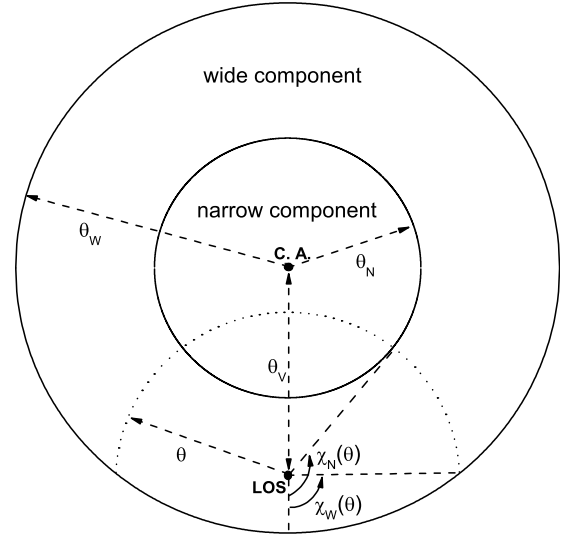,width=\columnwidth} \caption{Projection
  of a two-component jet. The inner cone is the narrow component,
  and the outer hollow cone is the wide component. The viewing angle between the jet axis (C.A.)
  and the line of sight (LOS) is $\theta_{V}$. Also shown are the apertures relative to the origin
  of the GRB jet, with $\theta_{N}$ of the narrow component and $\theta_{W}$ of the wide component.
  An arc centered to LOS with aperture $\theta$ (dotted line) is divided into three segments, with
  the segment ($\chi_{N}$,$2\pi-\chi_{N}$) in the narrow component and the segments ($\chi_{W}$,$\chi_{N}$)
  and ($2\pi-\chi_{N}$,$2\pi-\chi_{W}$) in the wide component. }
\end{figure}

The polarization of emission from a point-like region
averaged by the distribution of the post-shock tangled magnetic
field is (Sari 1999; Gruzinov 1999)
\begin{equation}
\Pi(\gamma,\theta)=\Pi_0\delta^{2}(\gamma,\theta)\sin^{2}\theta\frac{{\xi^{2}-1}}{{2\xi^{2}+(1-\xi^{2})\delta^{2}(\gamma,\theta)\sin^{2}\theta}},
\end{equation}
where $\theta$ is the angle between the velocity vector of the
emitting point and the line of sight,
$\delta(\gamma,\theta)=\displaystyle\frac{1}{\gamma(1-\beta\cos\theta)}$
is the Doppler boosting factor. The parameter $\xi^2=\langle
B_{\perp}^2\rangle/2\langle B_{\parallel}^2\rangle$ denotes the
level of anisotropy of the magnetic field distribution, where
$B_{\perp}$ and $B_{\parallel}$ are magnetic field components
perpendicular and parallel to the normal of the shock plane. The
triangular brackets mean that the quantity inside is averaged over
the solid angles. $\Pi_0$ is the linear polarization degree of
synchrotron photons emitted by the electrons in an ordered
magnetic field. For an isotropic power-law distribution of
electrons with index $p$, the approximate $\Pi_0=(p+1)/(p+7/3)$
holds true for a wide range of frequencies. The exact dependence
of $\Pi_0$ on the frequency is given in the Appendix (see Figure
A1). Using the above approximate expression for $\Pi_0$ actually
leads to neglectable errors in the final results. We therefore
adopt the usual approximate value of $\Pi_0$ in the following
calculations.

To determine the total polarization of a GRB jet, we need to
integrate the Stokes parameters over the jet surface. From Figure
1, by virtue of spherical geometry we have
\begin{equation}
\chi_{i}(\theta)=\left\{ {\begin{array}{c}
   \phantom{sssss} \pi\Theta(\theta_V-\theta_j) \phantom{ssssssss},  {\theta<\theta_{-}}, \\
   {\pi-\cos^{-1}(\displaystyle\frac{{\cos\theta_j-\cos\theta\cos\theta_V}}{{\sin\theta\sin\theta_V}})},\phantom{}{\theta_{-}<\theta<\theta_{+}},\\
   \phantom{ssssssssss} \pi \phantom{sssssssssssss},{\theta_{+}<\theta}, \\
\end{array}} \right.
\end{equation}
where $\theta_V$ is the observer's viewing angle with respect to
the jet axis, $\theta_{-}=\mid\theta_j-\theta_V\mid$, and
$\theta_{+}=\theta_j+\theta_V$. $\Theta(x)$ is the heaviside step
function with $\Theta(x>0)=1$ and $\Theta(x<0)=0$. Note that
equation (15) is valid for both components, i.e.
$\chi_{i}=\chi_{N}$ when $\theta_j=\theta_{N}$, and
$\chi_{i}=\chi_{W}$ when $\theta_j=\theta_{W}$. Ghisellini \&
Lazzati (1999) have given an approximate expression of $\chi_{i}$
($\psi_1$ in their work) for the intermediate case. Integrating
over each component of the jet, we get the Stokes parameters as
\begin{eqnarray}
{Q}_{i}(t,\nu)=\frac{{1+z}}{{D_{\rm{L}}^{2}}}&&\int_{0}^{\theta_{V}+\theta_{W}}{\delta^{3}(\gamma_{i},\theta)I'_{\nu',i}}\Pi(\gamma_{i},\theta)R_{i}^{2}\sin\theta d\theta \nonumber\\
                 & &\times\int_{\chi_{i,min}}^{\chi_{i,max}}{\cos2\varphi d\varphi},
\end{eqnarray}
\begin{eqnarray}
{I}_{i}(t,\nu)=\frac{{1+z}}{{D_{\rm{L}}^{2}}}&&\int_{0}^{\theta_{V}+\theta_{W}}{\delta^{3}(\gamma_{i},\theta)I'_{\nu',i}}R_{i}^{2}\sin\theta d\theta \nonumber\\
                 & &\times\int_{\chi_{i,min}}^{\chi_{i,max}}{d\varphi},
\end{eqnarray}
where $i=W$ or $N$, $\nu'=(1+z)\delta^{-1}(\gamma_{i},\theta)\nu$
is the frequency in the co-moving frame. For the narrow component,
$\chi_{N,min}=\chi_{N}(\theta)$ and
$\chi_{N,max}=2\pi-\chi_{N}(\theta)$. For the wide component, the
azimuthal integral is from $\chi_{W,min}=\chi_{W}(\theta)$ to
$\chi_{W,max}=\chi_{N,0}(\theta)$, and again from
$\chi_{W,min}=2\pi-\chi_{N,0}(\theta)$ to
$\chi_{W,max}=2\pi-\chi_{W}(\theta)$. Here $\chi_{N,0}(\theta)$ is
calculated by inserting the inner half-opening angle of the wide
hollow cone $\theta_{N,0}$ into equation (15), where
$\theta_{N,0}$ is the initial aperture of the narrow component.
Note that Stokes parameter $U_{N}=U_{W}=0$ can be easily derived
from the symmetry. The Stokes parameters for the whole jet are
therefore calculated by summation of the values of these two
components,
\begin{eqnarray}
{Q}_{tot}(t,\nu)=&&\frac{{1+z}}{{D_{\rm{L}}^{2}}}\int_{0}^{\theta_{V}+\theta_{W}}\sin\theta d\theta\times \nonumber\\
                  \{&&{\delta^{3}(\gamma_{W},\theta)I'_{\nu',W}}\Pi(\gamma_{W},\theta)R_{W}^{2}(\sin2\chi_{N,0}-\sin2\chi_{W})\nonumber\\
                &&-\delta^{3}(\gamma_{N},\theta)I'_{\nu',N}\Pi(\gamma_{N},\theta)R_{N}^{2}\sin2\chi_{N}\},
\end{eqnarray}
\begin{eqnarray}
{Q}_{tot}(t,\nu)=&&\frac{{1+z}}{{D_{\rm{L}}^{2}}}\int_{0}^{\theta_{V}+\theta_{W}}\sin\theta d\theta\times\nonumber\\
                  \{&&{\delta^{3}(\gamma_{W},\theta)I'_{\nu',W}}\Pi(\gamma_{W},\theta)R_{W}^{2}(\sin2\chi_{N,0}-\sin2\chi_{W})\nonumber\\
                &&-\delta^{3}(\gamma_{N},\theta)I'_{\nu',N}\Pi(\gamma_{N},\theta)R_{N}^{2}\sin2\chi_{N}\},
\end{eqnarray}
\begin{eqnarray}
{I}_{tot}(t,\nu)=&&\frac{{1+z}}{{D_{\rm{L}}^{2}}}\int_{0}^{\theta_{V}+\theta_{W}}\{{\delta^{3}(\gamma_{W},\theta)I'_{\nu',W}}R_{W}^{2}(2\chi_{N,0}-2\chi_{W})\nonumber\\
                 &&+\delta^{3}(\gamma_{N},\theta)I'_{\nu',N}R_{N}^{2}(2\pi-2\chi_{N})\}\sin\theta d\theta
                .
\end{eqnarray}
The total polarization of the two component jet is thus given by
\begin{equation}
P_{tot}=\displaystyle\frac{Q_{tot}}{I_{tot}}.
\end{equation}
The corresponding polarization vector is either (anti-) parallel
or perpendicular to the vector from C.A. to LOS projected in the
sky (see Figure 1), depending on whether the magnetic field
parameter $\xi^2$ is larger than unity or not, as well as on which
direction the observer views and how the jet evolves (Ghisellini
\& Lazzati 1999; Sari 1999).

\subsection{Numerical Results}
Below we use several sets of physical parameters to illustrate the
temporal evolution of afterglow polarization of a two-component
jet. The main interesting parameter effects may arise from the
ratio of the isotropic-equivalent energy of the two components
$E_{W,iso}/E_{N,iso}$, and the ratio of their initial half-opening
angles $\theta_{W,0}/\theta_{N,0}$. We fix other parameters and
assume $E_{N,iso}=10^{53}$ erg, $\theta_{N,0}=0.05$, $n=1$
cm$^{-3}$, $\epsilon_{e}=0.1$, $\epsilon_{B}=0.01$, $p=2.2$,
$\xi^2=9$ and $\Pi_0=60\%$ in the following calculations. The
equal arrival time surface effect is neglected in this primary
research. The adopted parameters of the cosmology are
$\Omega_{m}=0.27$, $\Omega_{\Lambda}=0.73$, $H_0=71$ km s$^{-1}$
Mpc$^{-1}$, and the GRB is assumed to be located at $z=1.0$.

Figure 2 shows the light curves and polarization evolutions of GRB
optical afterglows ($\nu_{opt}=4.5\times 10^{14}$ Hz) from
two-component jets\footnotemark\footnotetext{To compare which
component dominates the light curve or polarization, we treat the
two components separately, and then add them together. In the
figures we only give the total flux density and polarization.},
which are assumed to evolve with no lateral expansions. The ratio
of the wide component aperture to that of the narrow component is
chosen to be $\theta_{W,0}/\theta_{N,0}=6$. The upper panel
corresponds to a relatively low isotropic energy ratio as
$E_{W,iso}/E_{N,iso}=0.1$, while the lower panel corresponds to a
high ratio as $E_{W,iso}/E_{N,iso}=0.3$. When the line of sight is
in the narrow component, i.e. $\theta_{V}<\theta_{N,0}$, the light
curve differs little between each other since the light mainly
comes from the narrow component. The light curve steepens from
$\sim t^{-0.9}$ to $t^{-1.6}$ at around $t_{j,N}\sim 0.2$ day,
which is shallow as expected for the non-lateral expansion jet
(\meszaros\ \& Rees 1999). The polarization degree is also
dominated by the narrow component, and changes from an initially
negative value to a later positive value, which means that the
position angle of polarization changes by $90^{\circ}$. After
reaching a maximum, the polarization decreases and approaches
zero. When the line of sight lies within the wide component, the
light curve flattens when the narrow component decelerates and
enters into the field of view. The degree of flattening is
determined by the viewing angle $\theta_{V}$ and the energy ratio
$E_{W,iso}/E_{N,iso}$. The smaller $\theta_{V}$ or
$E_{W,iso}/E_{N,iso}$ is, the more contribution is resulted from
the narrow component, as can be seen in the figure. Contrary to
the light curve, the total polarization follows almost the
evolution of polarization of the narrow component for a long time.
This is because the narrow component is off-viewed and the
increase of the point polarization is greater than the decrease of
the Doppler boosting when $\theta$ increases (equations 14 and
16). The only exception is when $\theta_{V}/\theta_{N,0}=5$, in
which case the polarization evolution is dominated by the wide
component. For the other cases, the polarization level is depleted
to a certain extent by the wide component. In the cases of
$\theta_{V}/\theta_{N,0}=1.2$ and $1.5$, there exist long periods
with nearly constant polarization levels and unchanged position
angles, which is the typical characteristics of off-viewed jets
(Granot et al. 2002; Rossi et al. 2004). In general, increasing
the contrast between $E_{W,iso}$ and $E_{N,iso}$ would decrease
the maximum of polarization degree and reduce the flattening or
bump by the central narrow component.

\begin{figure}
\epsfig{figure=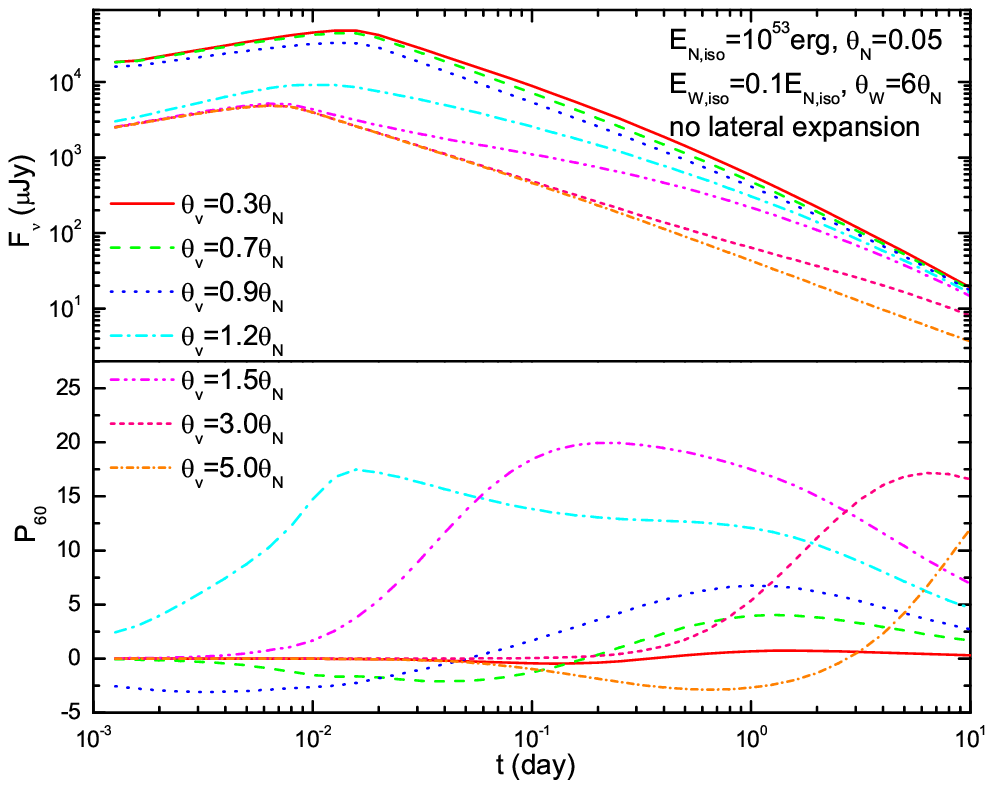,width=\columnwidth}
\epsfig{figure=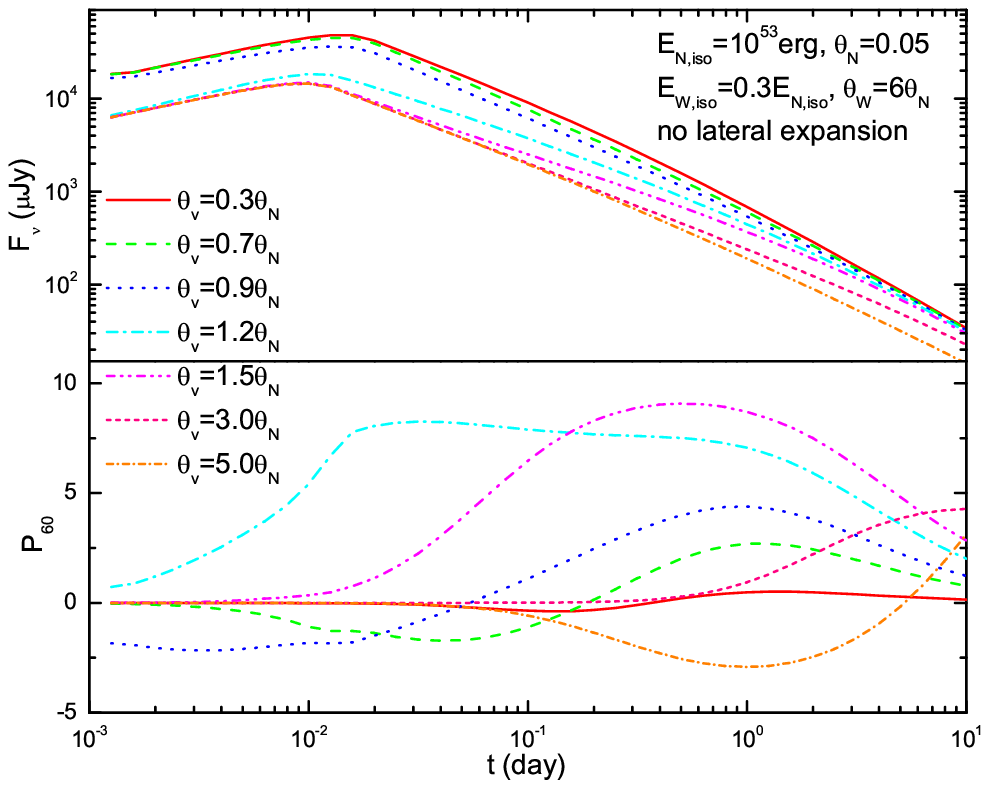,width=\columnwidth} \caption{Light curves
and polarization evolutions of GRB optical afterglows from
two-component jets. The jet is assumed to have no lateral
expansion. The contrasts between the wide and narrow component
parameters are $E_{W,iso}/E_{N,iso}=0.1$ (upper panel) or $0.3$
(lower panel), and $\theta_{W,0}/\theta_{N,0}=6$. Different lines
correspond to different viewing angles ($\theta_V$). }
\end{figure}

\begin{figure}
\epsfig{figure=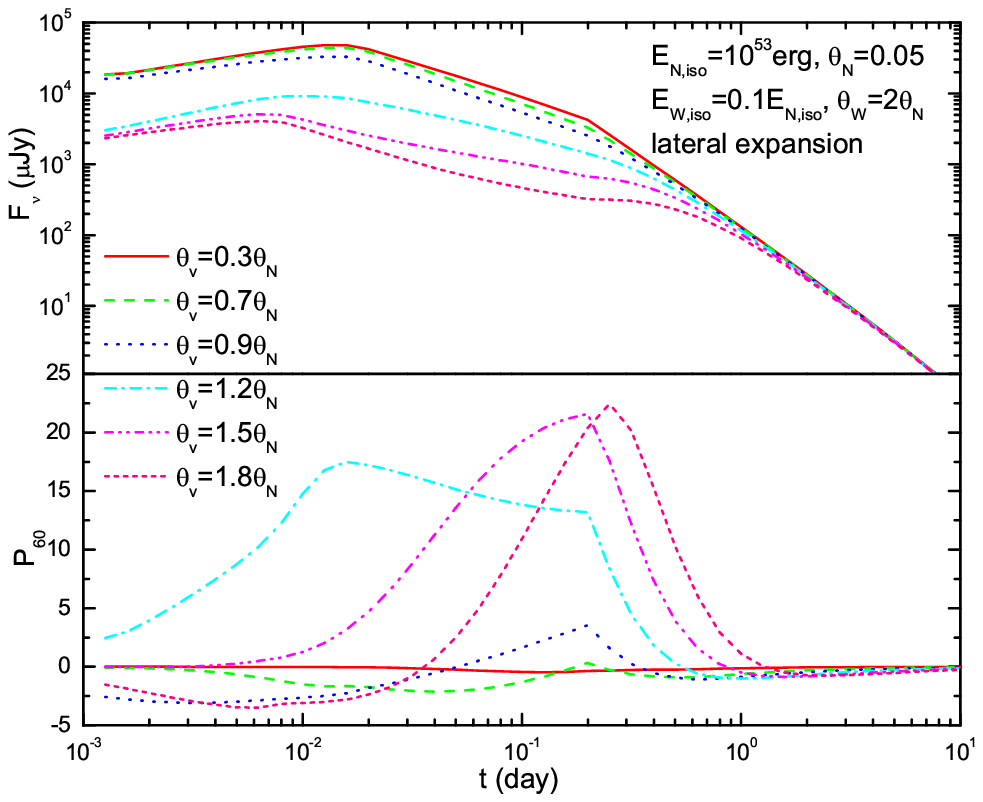,width=\columnwidth}
\epsfig{figure=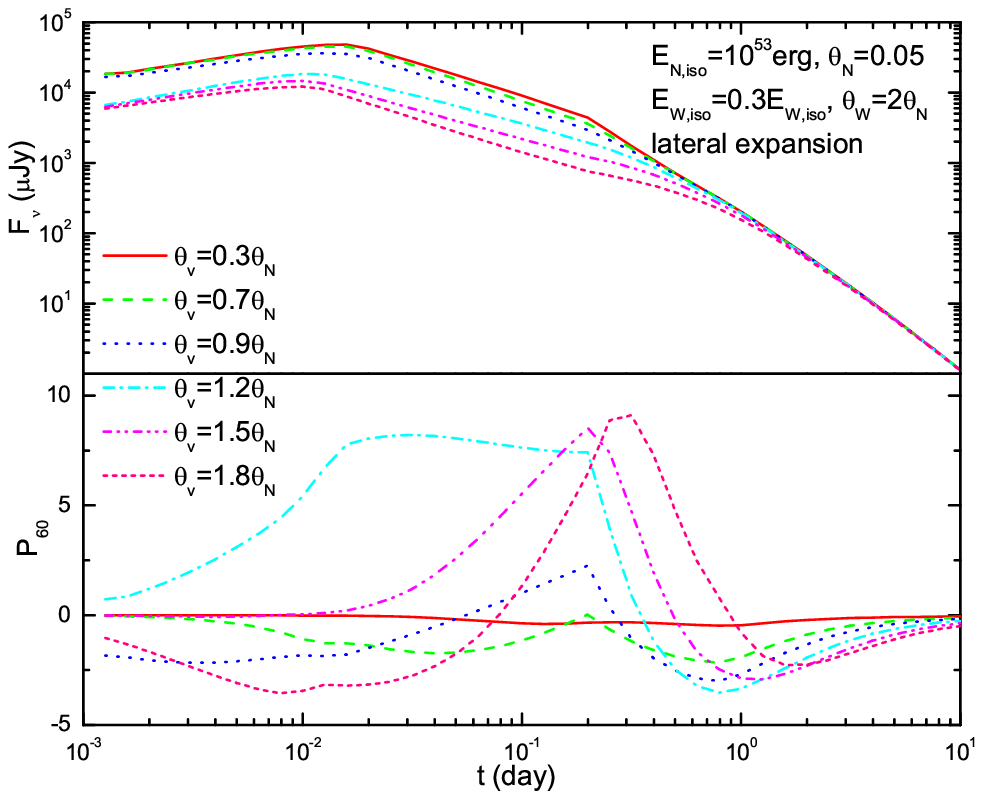,width=\columnwidth} \caption{Light curves
and polarization evolutions of GRB optical afterglows from
two-component jets. The jet is assumed to expand laterally at the
speed of light. The contrasts between the wide and narrow
component parameters are $E_{W,iso}/E_{N,iso}=0.1$ (upper panel)
or $0.3$ (lower panel), and $\theta_{W,0}/\theta_{N,0}=2$.
Different lines correspond to different viewing angles
($\theta_V$). }
\end{figure}

\begin{figure}
\epsfig{figure=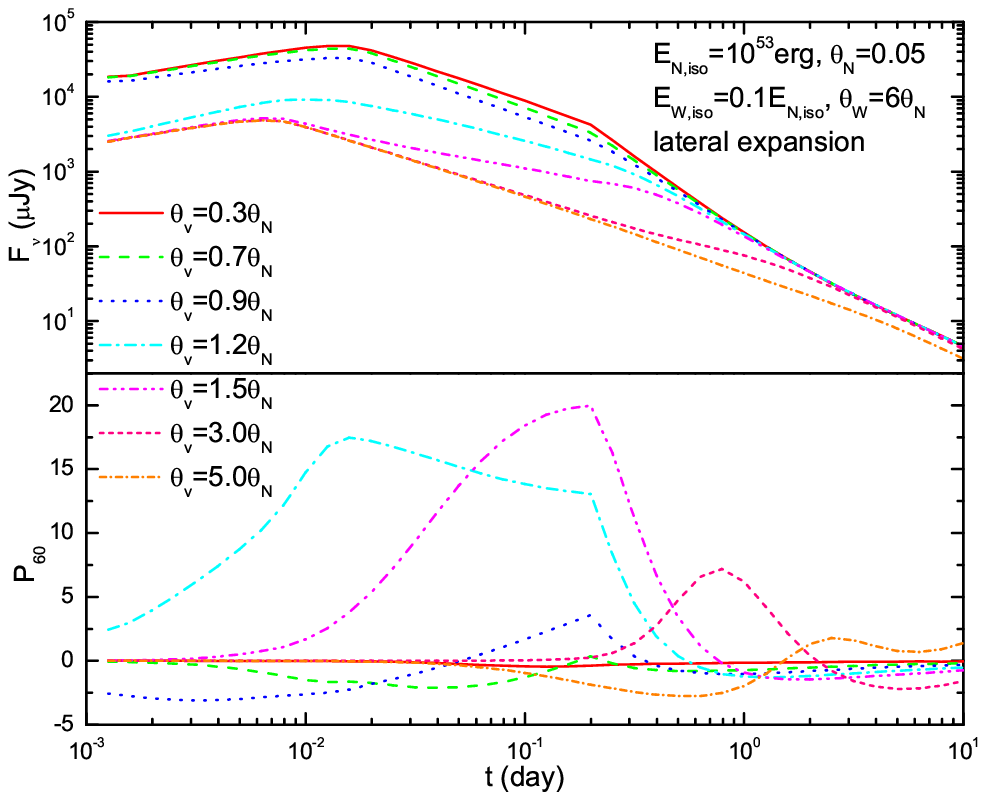,width=\columnwidth}
\epsfig{figure=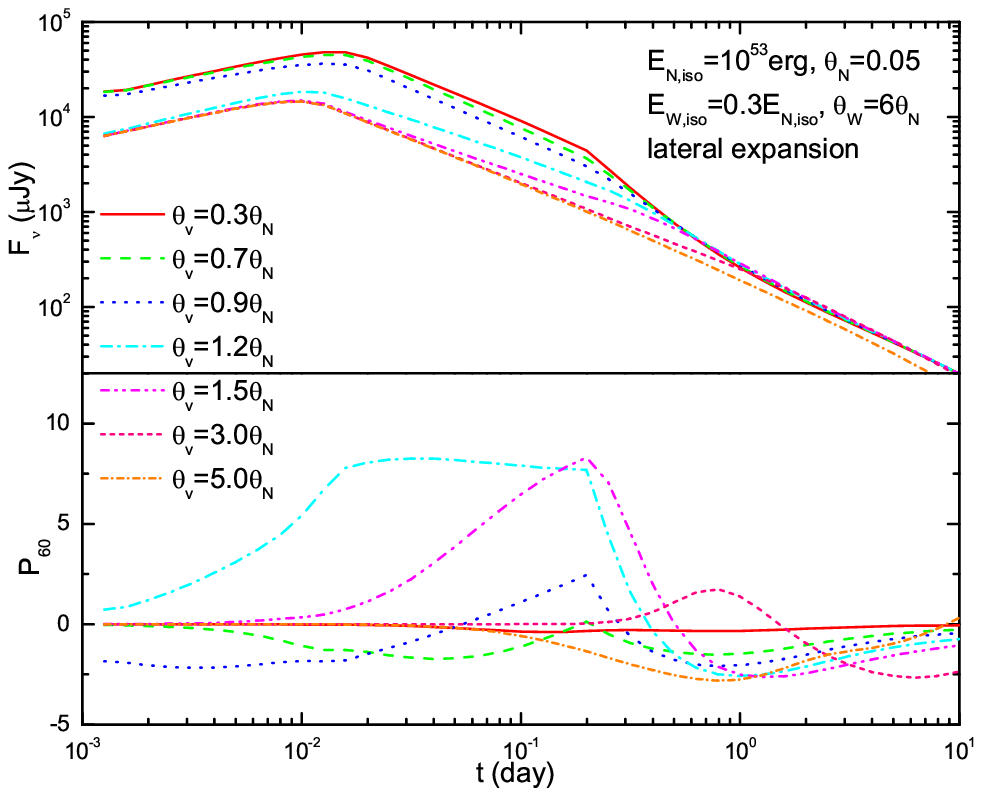,width=\columnwidth} \caption{Same as
Figure 3 except for $\theta_{W,0}/\theta_{N,0}=6$ now. }
\end{figure}

Figures 3 and 4 depict the light curves and polarization
evolutions of GRB optical afterglows from two-component jets with
lateral expansions. When $\theta_{V}<\theta_{N,0}$, the light
curve is dominated by the narrow component and experience a
steepening from $t^{-0.9}$ to $t^{-2.2}$ at $t_{j,N}\approx 0.2$
day. The light curve flattens at late times when the wide
component begins to dominate over the narrow one for
$\theta_{W,0}/\theta_{N,0}=6$, in which case the jet break of the
wide component occurs at rather late times, $t_{j,W}>10$
days\footnotemark\footnotetext{There is another possibility to
cause a late time flattening, in which case the jet (single
component) expanded sideways and becomes spherical and
non-relativistic (Huang \& Cheng 2003).}. The sign of polarization
changes twice for $\theta_{V}<\theta_{N,0}$. In most cases, except
when $\theta_{V}$ is very close to $\theta_{W,0}$, the
polarization curves have one peak at around $t_{j,N}$. These are
typical for jets with lateral expansion (Sari 1999). For
$\theta_{V}>\theta_{N,0}$, the polarization is mainly dominated by
the narrow component before $t_{j,N}$. Exceptions are
$\theta_{V}=1.8\theta_{N,0}$ in the case of
$\theta_{W,0}=2\theta_{N,0}$ and $\theta_{V}=5\theta_{N,0}$ in the
case of $\theta_{W,0}=6\theta_{N,0}$, where the polarization is
dominated by the wide component during the whole evolution. After
the jet break the asymmetry of the narrow component diminishes due
to sideways expansion, the polarization begins to be determined by
the wide component, almost regardless of whether $\theta_{V}$ is
larger than $\theta_{N,0}$ or not. In the figures we can see that
the absolute polarization level at $t>t_{j,N}$ increases with
$E_{W,iso}/E_{N,iso}$. As shown in the figures, the light curves
in cases of $\theta_{V}>\theta_{N,0}$ are influenced by the narrow
component through a bump or flattening. When the ratio
$\theta_{W,0}/\theta_{N,0}$ or especially $E_{W,iso}/E_{N,iso}$
increases, the influence of the narrow component to the light
curve is reduced, and the maximum polarization level also
decreases.

There are mainly two differences between the scenario without
lateral expansion and that with lateral expansion. The first is
the temporal behavior of light curve after the jet break time
$t_{j}$. The light-curve steepening of a non-lateral expansion jet
around $t_{j}$ is relatively shallow, while the steepening of a
lateral expansion jet is large. The second is the width of the
peak in the temporal polarization curve. A jet with lateral
expansion has a more narrow polarization peak than a jet without
lateral expansion, as can be seen by comparing Figure 2 with
Figure 3. Rossi et al. (2004) have investigated the polarization
evolutions of structured jets, with both the power-law and
Gaussian distributions. They found that the position angle of
polarization is not changed in time for these structured jets. A
two-component without lateral expansion also has an unchanged
position angle when $\theta_{N,0}<\theta_{V}<\theta_{W,0}$ . The
similarities between such two-component jets and those with power
law or Gaussian structure are that they all have no or little
lateral expansions and they are more energetic near the jet axis.

\subsection{The case of GRB $020813$}
The bright long duration GRB $020813$ detected by \emph{HETE-II}
spacecraft had been well-observed for evolutions of both optical
flux density and polarization. The redshift of this burst is
$z=1.255$, by identifying the [O II] $\lambda3727$ line of the
host galaxy (Barth et al. 2003). The light curve of GRB $020813$
afterglow is the smoothest one among GRBs localized so far
(Laursen \& Stanek 2003). The optical afterglow has a jet break
around $0.33$ - $0.88$ day since the main burst (Gorosabel et al.
2004; see also Covino et al. 2003b, Li et al. 2003). The temporal
index before the break ($\alpha_1$) is $\sim -0.55$ --- $\sim
-0.78$, and the index after the break ($\alpha_2$) is $\sim -1.44$
--- $-1.75$. Barth et al. (2003) have obtained the afterglow
spectrum in optical band, and found that the spectral index is
$\beta=1.06\pm 0.01$. Assuming that the spectrum is fast cooling,
the inferred power law index of electron distribution is
$p=2.12\pm0.02$. The spectropolarimetric observations had been
made by Keck between $0.19-0.33$ days and by VLT between
$0.88-4.05$ days (Barth et al. 2003; Gorosabel et al. 2004). The
degree of polarization decreases from about $3\%$ to less than
$1\%$, while the position angle varies slowly and can be regarded
as constant. Both observations showed that the majority of these
detected polarizations are intrinsic and not effected by the
line-of-sight dust. Therefore, the afterglow of GRB $020813$
provides us an ideal sample to investigate the structure of GRB
jets (Lazzati et al. 2004).

\begin{figure}
\epsfig{figure=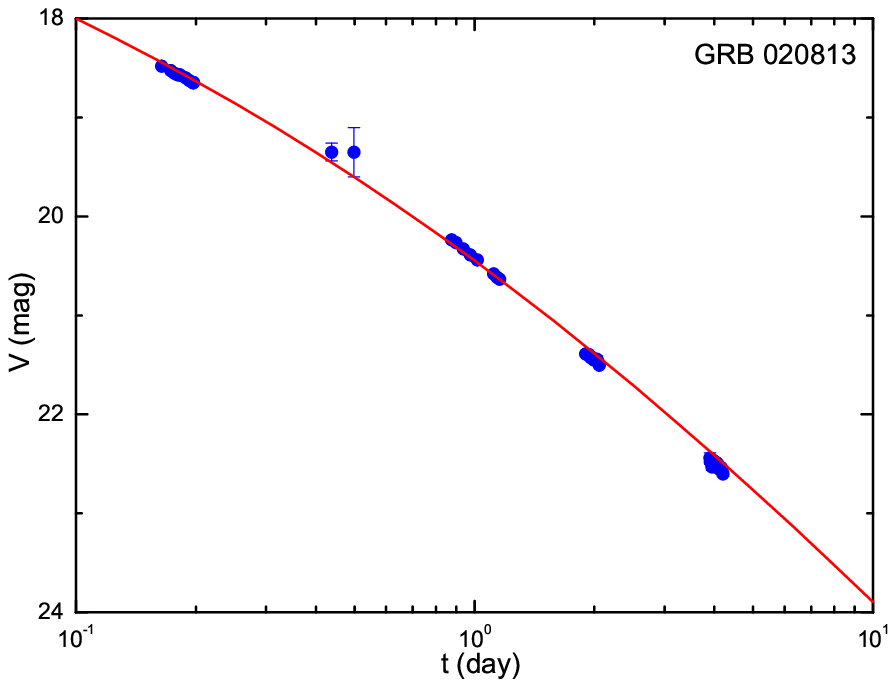,width=\columnwidth}
\epsfig{figure=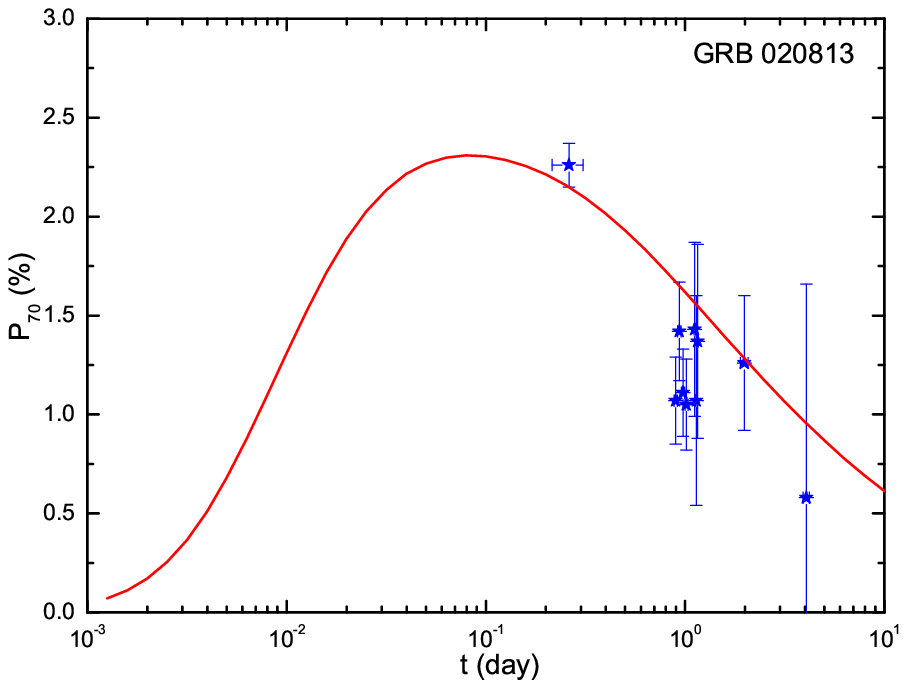,width=\columnwidth} \caption{Fitting to
the GRB 020813 V-band afterglow light curve (upper panel) and its
polarization evolution (lower panel). The Galactic extinction in
the V-band $A_{V}=0.42$ is adopted (Covino et al. 2003b). }
\end{figure}

We have fitted the V-band light curve and polarization evolution
of GRB $020813$ afterglow in the context of the two-component jet
model. The data of light curve and polarizations are taken from
Gorosabel et al. (2004), in which they included and averaged the
polarization observations made by Barth et al. (2003). We assume
that the position angle of polarization is unchanged, at least
within the time scale we concerned. The lateral spreading jet
scenario can be excluded for GRB $020813$ because the observed
temporal index $\alpha_2$ is larger than the inferred $-p$. Even
adopting a harder electron distribution with $1<p<2$, the
predicted $\alpha_2=-(p+6)/4$ is smaller than $-1.75$ (Dai \&
Cheng 2001). However, a non-lateral expansion jet running into the
ISM is able to reproduce the observed afterglow light curve. In
this case the theoretical $\Delta\alpha=\alpha_1-\alpha_2=0.75$ is
consistent with the observed value, ranging from
$\Delta\alpha_{obs,V}=0.50\pm0.17$ (Covino et al. 2003b) to
$\Delta\alpha_{obs,V}=1.20\pm0.17$ (Gorosabel et al. 2004). The
less constrained $\Delta\alpha_{obs,V}$ is due to the less
constrained jet break sharpness. Since the detected position angle
of polarization did not change around the jet break time
$t_{break,V}=0.33-0.88$ days, we expect the line of sight is
located within the wide component (e.g., see Figure 2 for the case
of $\theta_{V}=1.2\theta_{N,0}$). However, by analyzing the
polarization evolution, the viewing angle should be close to the
narrow component. Thus the wide component is responsible for the
main burst and early afterglow, while the narrow component is for
the whole evolution of the polarization and the late (mid) time
afterglow. The fittings to the GRB $020813$ V-band afterglow
including the light curve and polarization evolution are shown in
Figure 5, where parameters of the two-component jet are
$E_{N,iso}=1.2\times10^{54}$ erg, $\theta_N=0.02$,
$E_{W,iso}=0.2E_{N,iso}$, and $\theta_W=3.5\theta_N$. The jet is
viewed at $\theta_V=1.45\theta_N$. The shock parameters are
$\epsilon_{e}=0.05$, $\epsilon_{B}=5\times 10^{-5}$, $p=2.15$ and
$\xi^2=1.25$. The ISM density is $n=1.0$ cm$^{-3}$. The relatively
low $\xi^{2}$ was also indicated by Barth et al. (2003). Granot \&
K\"{o}nigl (2003) proposed that the magnetic field configuration
may be isotropized by the post-shock turbulence, and $\xi^2$ we
observed is the averaged value over the shell width.

Due to the sparse detections of polarizations, especially at
$t\sim 0.5$ day, Barth et al. (2003) argued that it is still
possible to accommodate the lateral spreading jet model for GRB
$020813$. But a structured jet, with the center to be brighter or
more energetic while the wings to be dimmer or less energetic,
seems more natural as compared with other models according to
polarization observations (Lazzati et al. 2004). However, we
should also bear in mind that the details of the structures, i.e.,
whether its profile is power-law, Gaussian, or two-component,
still cannot be definitely determined currently.

\section{Conclusions and Discussion}
In this paper we have investigated the polarization property of a
two-component GRB jet. The inner component is narrow and more
energetic while the outer one is wide and less energetic. The
hydrodynamic evolution of each component follows the relativistic
Blandford-McKee scaling law in the ISM case. The shock physics is
assumed to be the same for these two components. We include the
inverse Compton scattering in the calculations of afterglow light
curves. The major effect of inverse Compton scattering is to
further reduce the cooling Lorentz factor of electrons. Since the
typical energy of inverse Compton scattered photons lies above the
X-ray band, and the synchrotron-self absorption operates in the
radio afterglow, we considered only the synchrotron origin of
optical polarization.

The resulting optical light curves and polarization evolutions depend strongly on the ratios of the
intrinsic parameters of the wide component to those of the narrow one (i.e. $E_{W,iso}/E_{N,iso}$
and $\theta_{W,0}/\theta_{N,0}$; see also Peng, K\"{o}nigl, $\&$ Granot 2004, in which they
investigated the light curves of GRB afterglows within the two-component jet model in analytical
way), on the different assumptions of the lateral expansion, and on the observer's viewing angle as
well. Two scenarios of different lateral expansions are studied. One is to assume that the jet has
no lateral expansion, while another is to assume the jet expands laterally at the speed of light.
In both cases we find that within a wide range of the viewing angle the polarization is dominated
by the narrow component for a long period. For an observer with the line of sight located in the
wide component, the position angle of polarization will not change if the jet has no lateral
expansion. This is because the geometric asymmetry of the narrow component does not change with
respect to the line of sight. On the other hand, the position angle will change by 90 degree for a
laterally expanding jet since the jet geometry evolves from an asymmetrical one to a symmetrical
one, with the transition taking place around $t_{j}$ (Ghisellini \& Lazzati 1999; Sari 1999).
Although the two-component jet model has been proposed to explain some peculiar afterglow light
curves (Frail et al. 2000; Berger et al. 2003; Huang et al. 2004), it does not imply necessarily
that the light curve of a two-component jet must be peculiar. When the line of sight lies within
one component, another component may cause a flattening of the light curve at late times, not
necessarily a bump, depending on the contrast of the parameters between these two components (see
also Huang et al. 2004, where the equal arrival time surface effect was included).

We find that the two-component jet model with the assumption of no
lateral expansion is able to explain the afterglow of GRB $020813$
through fittings to its optical light curve and polarization
evolution. The essence of this explanation is to assume the line
of sight is located in the wide component, which ensures a
constant position angle of polarization. However, as pointed by
Lazzati et al. (2004), any jet with a more energetic core and less
energetic wings will produce qualitatively similar polarization
curves. Due to the sparse polarization data of GRB $020813$
afterglow, especially at very early times and around the jet break
time $t_{j}\sim 0.5$ day, we cannot determine the actual structure of
GRB jets definitely. In the upcoming \emph{Swift} era, we expect that several
well-observed GRB afterglows with copious data of both magnitude and
polarizations will give us new clues about jet structures.

We thank the referee for his/her invaluable suggestions and
detailed comments which have led us to improve this paper
significantly. XFW would like to thank X. Y. Wang for his
encouragement. This work was supported by the National Natural
Science Foundation of China (grants 10233010, 10221001, 10003001,
and 10473023), the Ministry of Science and Technology of China
(NKBRSF G19990754), the Special Funds for Major State Basic
Research Projects, and the Foundation for the Authors of National
Excellent Doctoral Dissertations of P. R. China (Project No.
200125).

\begin{appendix}
\section{Evaluation of the synchrotron $\Pi_0$ in an ordered magnetic field}
The linear polarization of optically thin synchrotron radiation in
an ordered magnetic field is calculated by averaging over an
isotropic distribution of electrons (Longair 1994)
\begin{equation}
\Pi_0(\nu')=\frac{\int G(x)N(\gamma_{e})d\gamma_{e}}{\int
F(x)N(\gamma_{e})d\gamma_{e}},
\end{equation}
where $x=\nu'/\nu'_{\gamma_{e}}$ with
$\nu'_{\gamma_{e}}=\displaystyle\frac{3\gamma_{e}^{2}eB'}{4\pi
m_{e}c}$, $F(x)=x\int_{x}^{\infty}K_{5/3}(t)dt$ and $G(x)=x
K_{2/3}(x)$, where $K_{5/3}(x)$ and $K_{2/3}$ are the modified
Bessel functions. In the limit of $x\ll 1$,
$F(x)\approx2G(x)\approx\displaystyle\frac{4\pi}{\sqrt{3}\Gamma(1/3)}(\frac{x}{2})^{1/3}$,
while in the limit of $x\gg 1$, $F(x)\approx
G(x)\approx\displaystyle\sqrt{\frac{\pi}{2}}x^{1/2}e^{-x}$. The
electron distribution function, $\displaystyle
N(\gamma_{e})=\frac{dN}{d\gamma_{e}}$, determines the dependence
of $\Pi_0$ on the frequency $\nu'$.

\textit{Fast-cooling case} ($\gamma_{c}<\gamma_{m}$). The electron
distribution is approximated by a broken power-law, with
$N(\gamma_{e})\propto\gamma_{e}^{-2}$ for
$\gamma_{c}<\gamma_{e}<\gamma_{m}$ and
$N(\gamma_{e})\propto\gamma_{e}^{-p-1}$ for
$\gamma_{m}<\gamma_{e}<\gamma_{M}$, where $\gamma_{M}$ is the
maximum Lorentz factor of accelerated electrons. Rewriting
equation (A1), we get
\begin{equation}
\Pi_0(\nu')=\frac{\int_{x_{m}}^{x_{c}}G(x)x^{-1/2}dx+x_{m}^{(1-p)/2}\int_{x_{M}}^{x_{m}}G(x)x^{(p-2)/2}dx}
                 {\int_{x_{m}}^{x_{c}}F(x)x^{-1/2}dx+x_{m}^{(1-p)/2}\int_{x_{M}}^{x_{m}}F(x)x^{(p-2)/2}dx},
\end{equation}
where $x_{c}=\nu'/\nu'_{c}$, $x_{m}=\nu'/\nu'_{m}$ and
$x_{M}=\nu'/\nu'_{M}$. $\Pi_0$ approaches $1/2$ for
$\nu'\ll\nu'_{c}\ll\nu'_{m}$, $9/13$ for
$\nu'_{c}\ll\nu'\ll\nu'_{m}$, and $(p+2)/(p+10/3)$ for
$\nu'_{m}\ll\nu'\ll\nu'_{M}$.

\textit{Slow-cooling case} ($\gamma_{m}<\gamma_{c}$). The electron
distribution is approximated by a broken power-law, with
$N(\gamma_{e})\propto\gamma_{e}^{-p}$ for
$\gamma_{m}<\gamma_{e}<\gamma_{c}$ and
$N(\gamma_{e})\propto\gamma_{e}^{-p-1}$ for
$\gamma_{c}<\gamma_{e}<\gamma_{M}$. Equation (A1) can be deduced
to be
\begin{equation}
\Pi_0(\nu')=\frac{\int_{x_{c}}^{x_{m}}G(x)x^{(p-3)/2}dx+x_{c}^{-1/2}\int_{x_{M}}^{x_{c}}G(x)x^{(p-2)/2}dx}
                 {\int_{x_{c}}^{x_{m}}F(x)x^{(p-3)/2}dx+x_{c}^{-1/2}\int_{x_{M}}^{x_{c}}F(x)x^{(p-2)/2}dx},
\end{equation}
which approaches $1/2$ for $\nu'\ll\nu'_{m}\ll\nu'_{c}$,
$(p+1)/(p+7/3)$ for $\nu'_{m}\ll\nu'\ll\nu'_{c}$, and
$(p+2)/(p+10/3)$ for $\nu'_{c}\ll\nu'\ll\nu'_{M}$.

\begin{figure}
\epsfig{figure=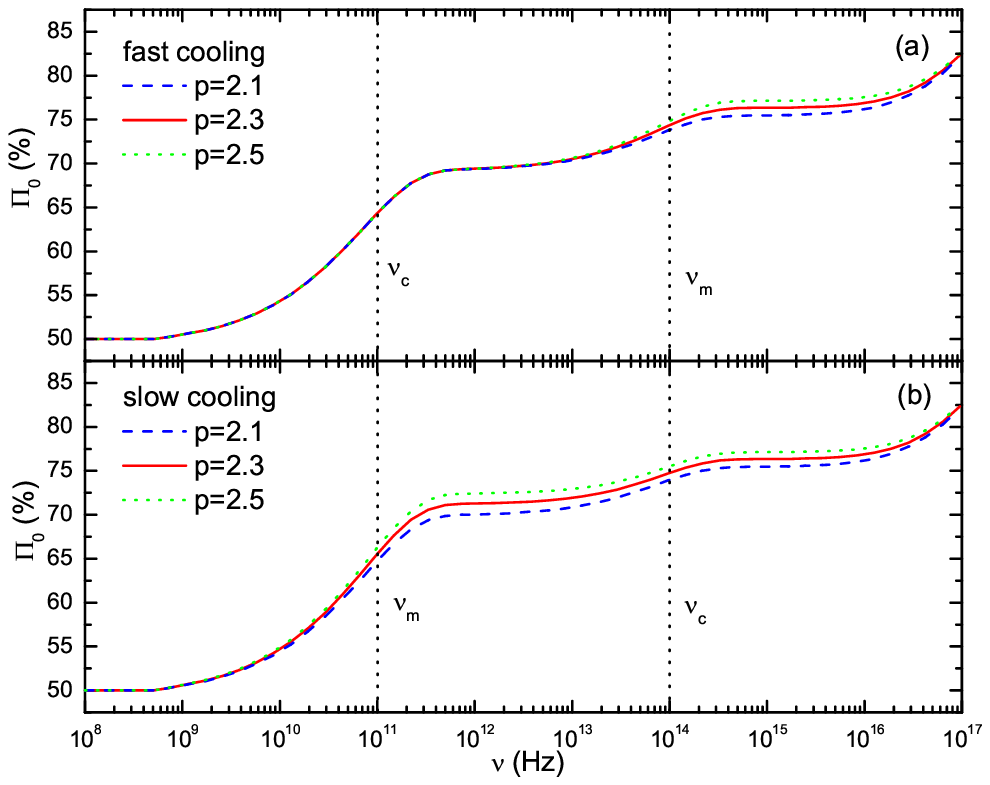,width=\columnwidth} \caption{The dependence
of $\Pi_0$ on the observed frequency $\nu$. Panel (a): the
fast-cooling case ($\nu_{c}<\nu_{m}$);
  Panel (b): the slow-cooling case ($\nu_{c}>\nu_{m}$). The dashed, solid and dotted lines correspond to
  $p=2.1$, $2.3$, and $2.5$ respectively. }
\end{figure}

The above approaches are consistent with those in Table 1 of
Granot (2003), in which the simple analytic expression for $\Pi_0$
as a function of the spectral slope for optically thin synchrotron
emission can be found. Figure A1 clearly shows how the
polarization degree $\Pi_0$ varies with the observed frequency
$\nu=\delta(\gamma,\theta)(1+z)^{-1}\nu'$. As can be seen, in both
fast and slow cooling cases, there are three platforms in each
polarization curve, which represent the three typical polarization
levels discussed above.

\end{appendix}

\end{document}